\documentclass[epj]{webofc}
\usepackage[varg]{txfonts}   
\usepackage{graphicx,color} 
\usepackage{bm}       
\usepackage{amsmath}  
\usepackage{amssymb}
\woctitle{21st International Conference on Few-Body Problems in Physics}
\usepackage[normalem]{ulem}

\begin{document}
\title{Direct Bethe-Salpeter solutions in Minkowski space}
\author{J.~Carbonell\inst{1}\fnsep\thanks{\email{carbonell@ipno.in2p3.fr}} \and
        V.A.~Karmanov\inst{2}\fnsep\thanks{\email{karmanov@sci.lebedev.ru}} 
}

\institute{Institut de Physique Nucl\'eaire,
Universit\'e Paris-Sud, IN2P3-CNRS, 91406 Orsay Cedex, France
\and
           Lebedev Physical Institute, Leninsky Prospekt 53, 119991
Moscow, Russia 
          }

\abstract{
 We review a method  to directly solve  the  Bethe-Salpeter equation in Minkowski space, both for bound and scattering states.
It is based on a proper treatment of the many singularities 
which appear in  the kernel and propagators. The off-mass shell scattering amplitude for spinless particles interacting by a one boson exchange was  computed for the first time.  Using our Minkowski space solutions  
for the initial (bound) and  final (scattering) states, we calculate elastic and transition (bound $\to$ scattering state) electromagnetic form factors.  The conservation of the transition electromagnetic current $J\cdot q=0$, verified numerically, confirms the validity of our solutions.
}
\maketitle

\section{Introduction}
\label{intro}

Bethe-Salpeter (BS) approach, formulated originally in the Minkowski space, provides a self-consistent  description of relativistic systems based on the field theory.  The main reason  for being interested  in the Minkowski solutions is the fact that the Euclidean ones are not able to provide some fundamental observables.
While the purely Euclidean solutions  can be used to obtain  the binding energy,  
the on-shell scattering observables (phase shifts) require already coupled equations  with some particular Minkowski amplitudes\cite{tjon,ck_bs_long},
As we have shown in ~\cite{ckm_ejpa,ck-trento}, Wick rotation is not directly applicable
for computing electromagnetic (e.m.) form factors when the BS amplitudes are known numerically.  
Finally, the computation of the off-shell BS scattering amplitude  -- mandatory for finding e.g. the transition  e.m.  form factor --  is possible only using a full Minkowski solution.

For bound states, the Minkowski solutions were found in \cite{Kusaka}, by representing the BS amplitude via Nakanishi integral \cite{nak63}.
A similar approach combined with the light-front projection  was proposed in \cite{bs1-2}. A modified formalism to the one developed in \cite{bs1-2} aimed to compute the scattering states was developed in \cite{fsv-2012}. It has been  tested for bound states \cite{FrePRD14} and, more recently, for  computing the -- zero-energy -- scattering lengths   \cite{fsv_scat_length}.

We present here the off- and on-shell solutions  the BS equation in Minkowski space found by  a direct method   \cite{ck_bs_long}
based on a proper treatment of the many singularities which appear in  the kernel and propagators.  
It is applicable  to  bound and scattering states, and valid in any energy domain. 

\section{BS amplitude and phase shifts}\label{sec-1}
We have first computed  the bound state solutions. The binding energies  thus obtained  coincide, within four-digit accuracy, with the ones calculated in our previous work \cite{bs1-2}.

For the scattering states, in c.m.-frame  $\vec{p}=0$, the BS off-shell partial wave amplitude $F^{off}$, for a given incident momentum $p_s$ (corresponding to given invariant mass of the scattering state $M'=2\sqrt{m^2+p_s^2}$)  depends on two variables  $k_0,k$: $F^{off}=F(k_0,k;p_s)$.
The BS  amplitude $F$ is factorized so that the strongest singularities are included in a known analytical prefactor, and the remaining regular part $f$ is expanded in a spline basis.  In this way, the BS equation is transformed in a numerically solvable form and the off-mass shell scattering amplitude for spinless particles interacting by a one boson exchange (with the mass $\mu$) was  computed for the first time. The detail of this transformation are given in \cite{ck_bs_long}.
\vspace{0.3cm}
\begin{figure}[h!] 
\begin{center}
\includegraphics[width=5.cm]{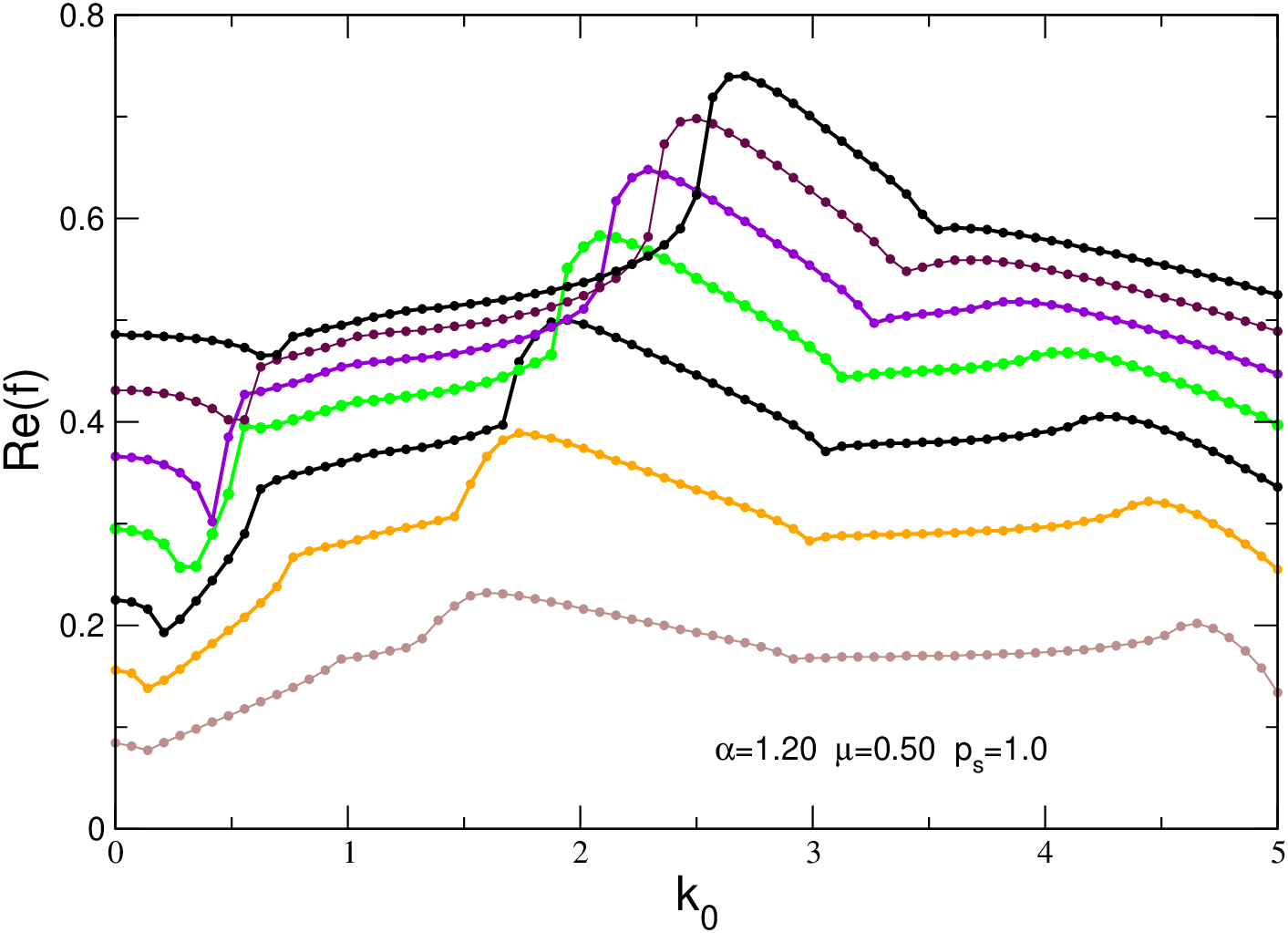}\quad
\includegraphics[width=5.cm]{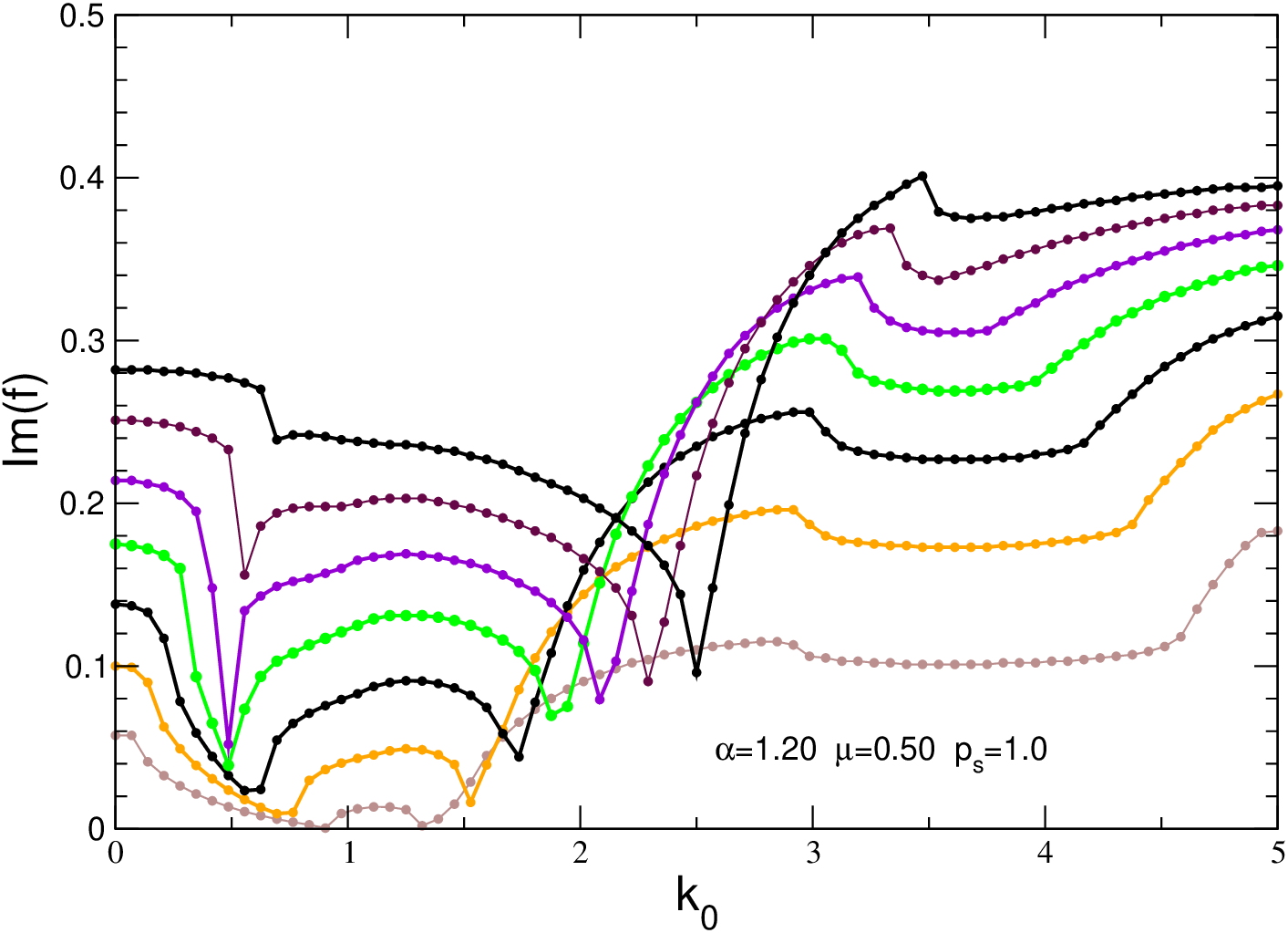}
\caption{Real (left panel) and imaginary (right panel) parts of the regular part  $f_0(k,k_0)$ of the off-shell scattering amplitude vs. $k_0$ for a few fixed values of $k$  and $\mu=0.5$ (in the units of mass $m$).}\label{FIG_f_k0}
\end{center}
\end{figure} 
\vspace{-.5cm}

As an example, we display in Fig. \ref{FIG_f_k0} the regular part  $f_0(k,k_0)$ of the S-wave off-shell scattering amplitude
as a function $k_0$ for different values of $k$, for the coupling constant $\alpha=g^2/(16\pi m^2)=1.2$.  As one can see, this function is no longer singular, although it has highly non-trivial behavior with several sharp structures and cusps both in its real and imaginary parts. 
When the incident momentum $p_s$ exceeds the threshold of the boson creation, the amplitude has a cusp and  the phase shift automatically obtains an imaginary part. 
Some of these sharp behaviors can be understood analytically. In these -- rare -- cases,  the numerical results are in close agreement with the analytical predictions giving
thus a strong confidence on the validity of our approach (see for instance Fig.  12 in Ref.~\cite{ck_bs_long}). 

The on-mass shell condition $k_1^2=k_2^2=m^2$, in the relative variables $k_0,k$, obtains the form $k_0=0,k=p_s$.  The corresponding on-shell amplitude  $F^{on}=F(k_0=0,k=p_s;p_s)=F(p_s)$ is the physical scattering amplitude related to the phase shift $\delta$ as:
$$
F^{on}=\frac{M'}{4ip_s}\left[\exp(2i\delta)-1\right]
$$
\begin{figure}[h!]
\centering
\includegraphics[width=5cm]{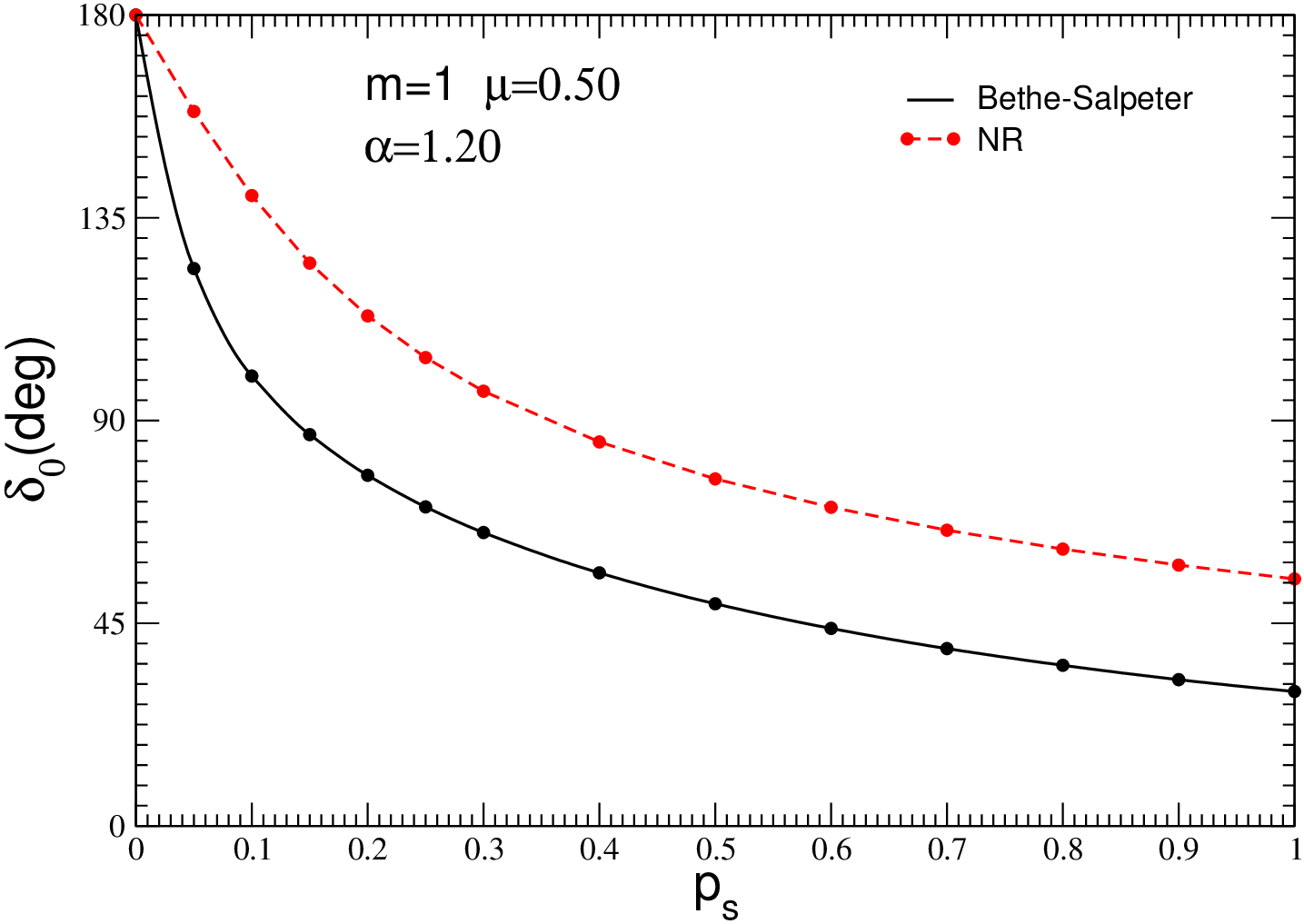} 
\quad
\includegraphics[width=5cm]{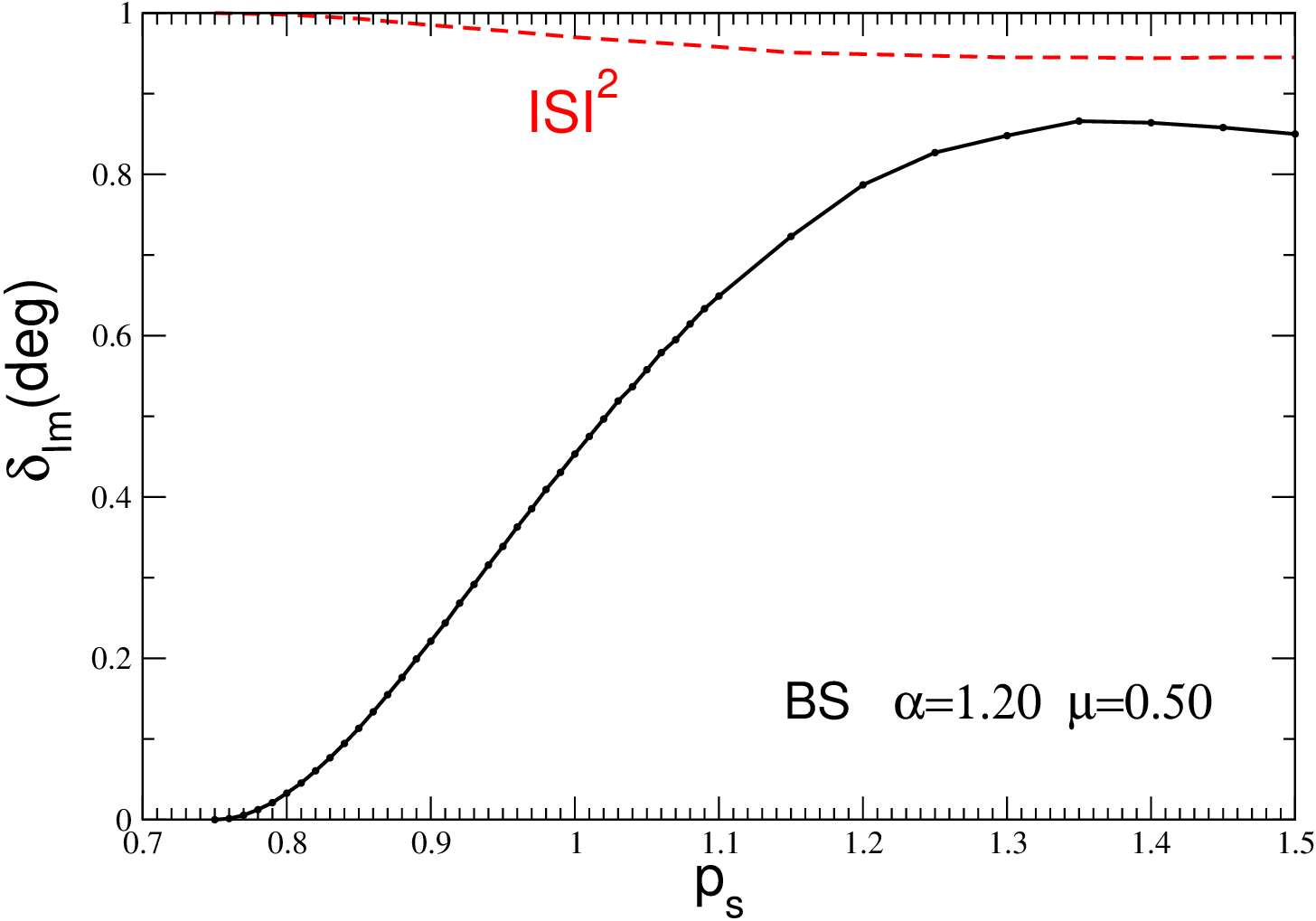}
\caption{Left panel: real part of the phase shift (degrees) for $\alpha=1.2$ and $\mu=0.50$ calculated  via BS equation (solid)  compared to the non-relativistic results (dashed). Right panel: imaginary part of the phase shift (degrees) calculated  via BS equation (solid) and the inelasticity parameter  $\eta=|S|^2=\exp[-2\mbox{Im}(\delta)]$ (dashed).}
\label{fig2}
\end{figure}

Figure \ref{fig2}  (left panel) shows the real part of the phase  shifts calculated via the BS equation (solid line) as a function of the scattering momentum $p_s$.
They are compared to the non-relativistic (NR)  values (dashed lines) provided by the Schr\"odinger equation with the Yukawa potential.
For the value of $\alpha=1.2$ there exists a bound state and, according to the Levinson theorem, the phase shift starts at 180$^\circ$.
One can see that the difference between relativistic and non-relativistic results is considerable even for relatively small incident momentum.

The right panel  shows the imaginary part of the phase shift. It appears  starting from the first inelastic meson-production threshold $p^{(1)}_s=0.75$ and in vicinity of $p_s=p^{(1)}_s$ displays the expected quadratic behavior  ${\rm Im}[\delta_0]\sim (p_s-p^{(1)}_s)^2$.
Simultaneously the modulus squared of the S-matrix (dashed line) starts differ from unity.
We reproduce the phase shifts found in \cite{tjon}.
\vspace{0.3cm}
\begin{figure}[hbtp]
\centering
\includegraphics[width=5cm]{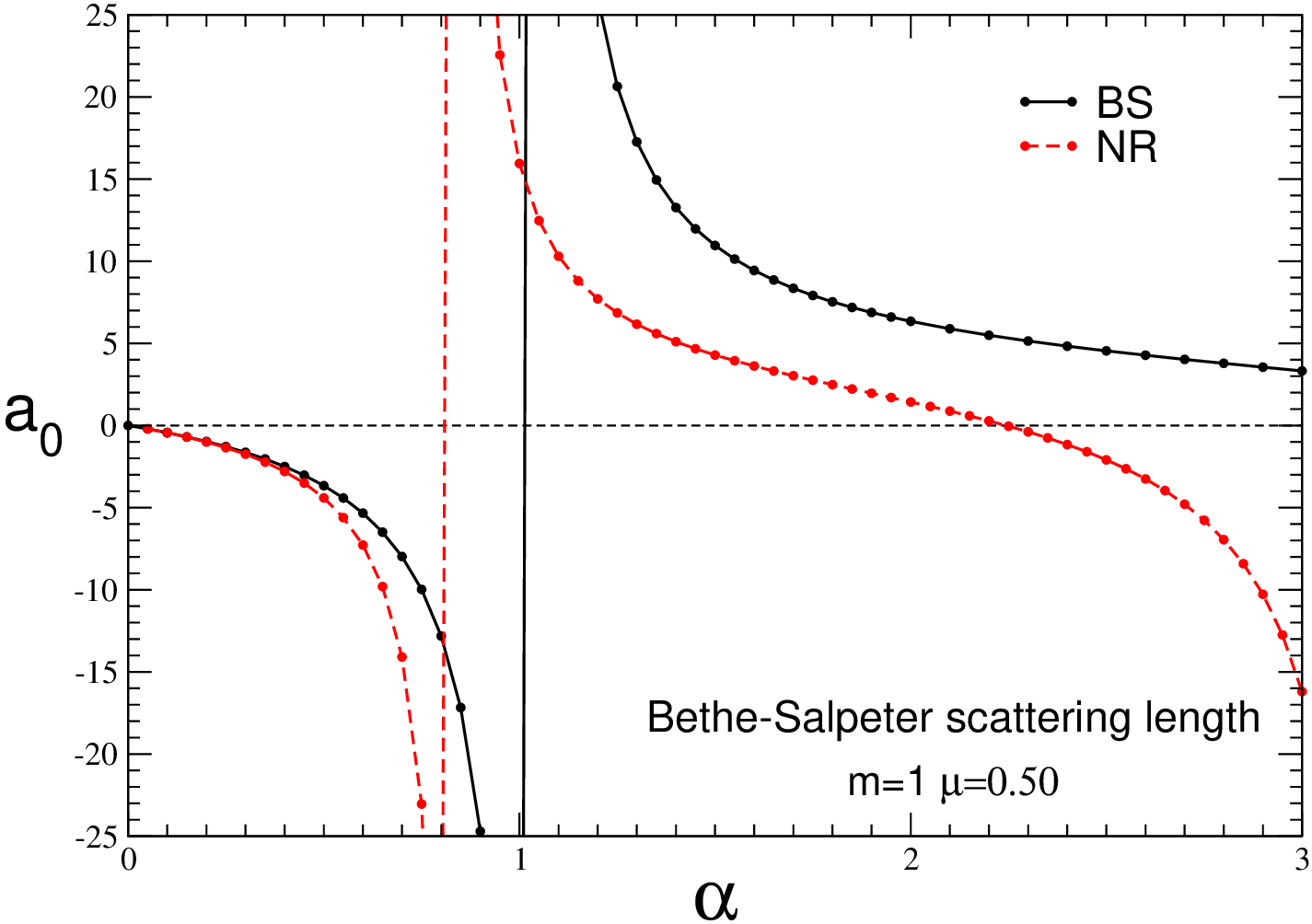}
\caption{BS scattering length $a_0$ versus  $\alpha$ (solid), compared to the non-relativistic results (dashed) for $\mu=0.5$.}\label{fig_a0}
\end{figure}
The low energy parameters were computed directly at $p_s=0$
and found to be consistent with a quadratic fit to the effective range function \mbox{$k\cot\delta_0(k)= -\frac{1}{a_0} + \frac{1}{2} r_0 k^2 $.}
The BS scattering length  $a_0$ as a function of the coupling constant $\alpha$ is given in Fig. \ref{fig_a0} for  $\mu=0.50$.
It is compared to the non-relativistic (NR)  values.
The singularities correspond to the appearance of the first bound state at $\alpha_0=1.02$ for BS and  $\alpha_0=0.840$ for NR.
One can see that the differences between relativistic and non-relativistic treatments is substantially large even in processes involving zero energy,  especially in presence of a bound state.

Our results for the scattering length coincide with good precision with ones found in \cite{fsv_scat_length} by a different method. Some discrepancy for small $\mu$, indicated in  \cite{fsv_scat_length}, disappeared after taking the limit $p_s\to 0$ in our equations analytically.

The method can be easily generalized to fermions, since   boson and  fermion propagators have  same  singularities.
This opens the way for studying the NN system   in the BS framework.

\section{E.M. form factors}

Using our Minkowski space solutions  for the initial (bound) and  final (scattering) states, we calculate the elastic form factor  \cite{ckm_ejpa}
 and the transition one for electrodisintegration of a two-body system \cite{ck_fft}.
\begin{figure}[h!]
\begin{center}
\includegraphics[width=5cm]{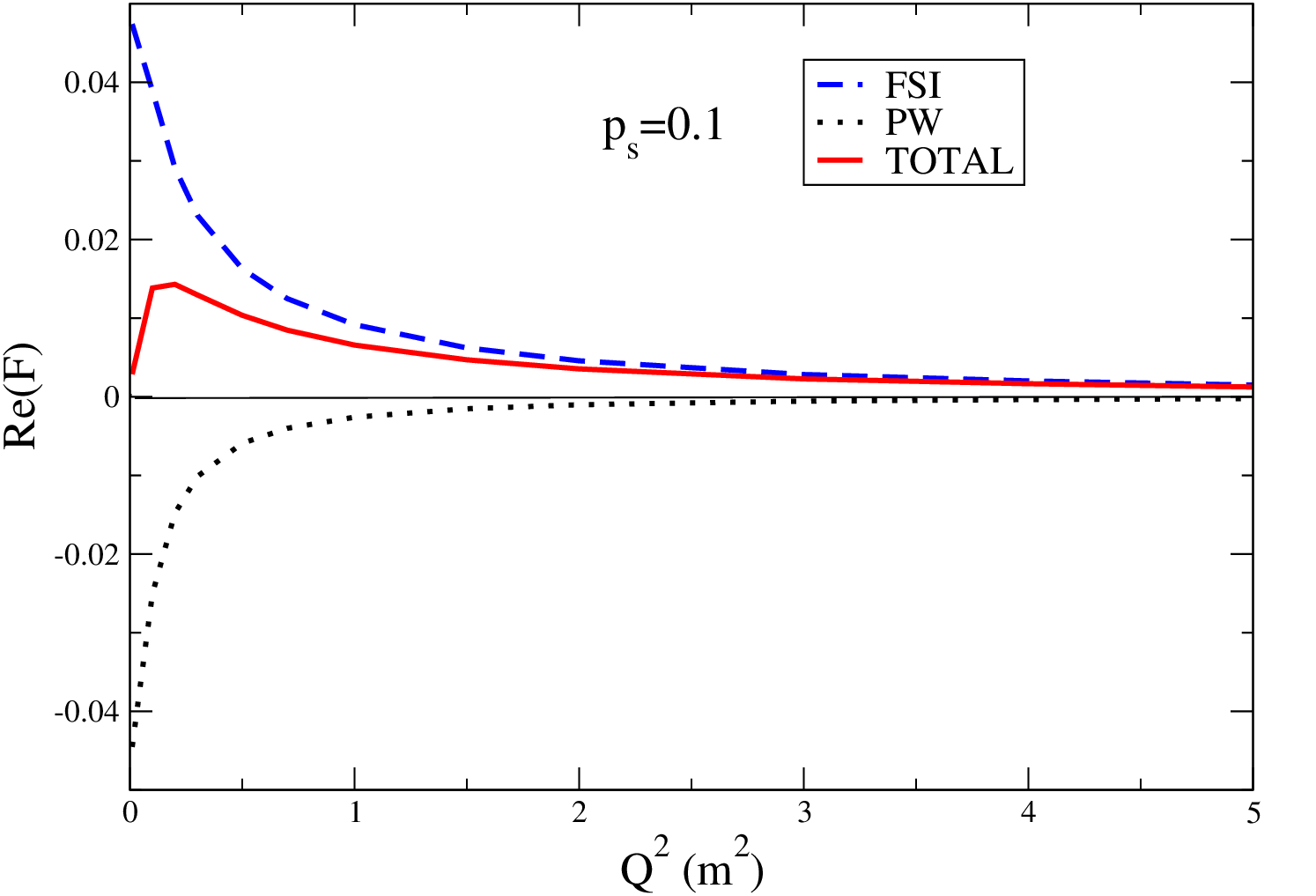}
\quad
\includegraphics[width=5cm]{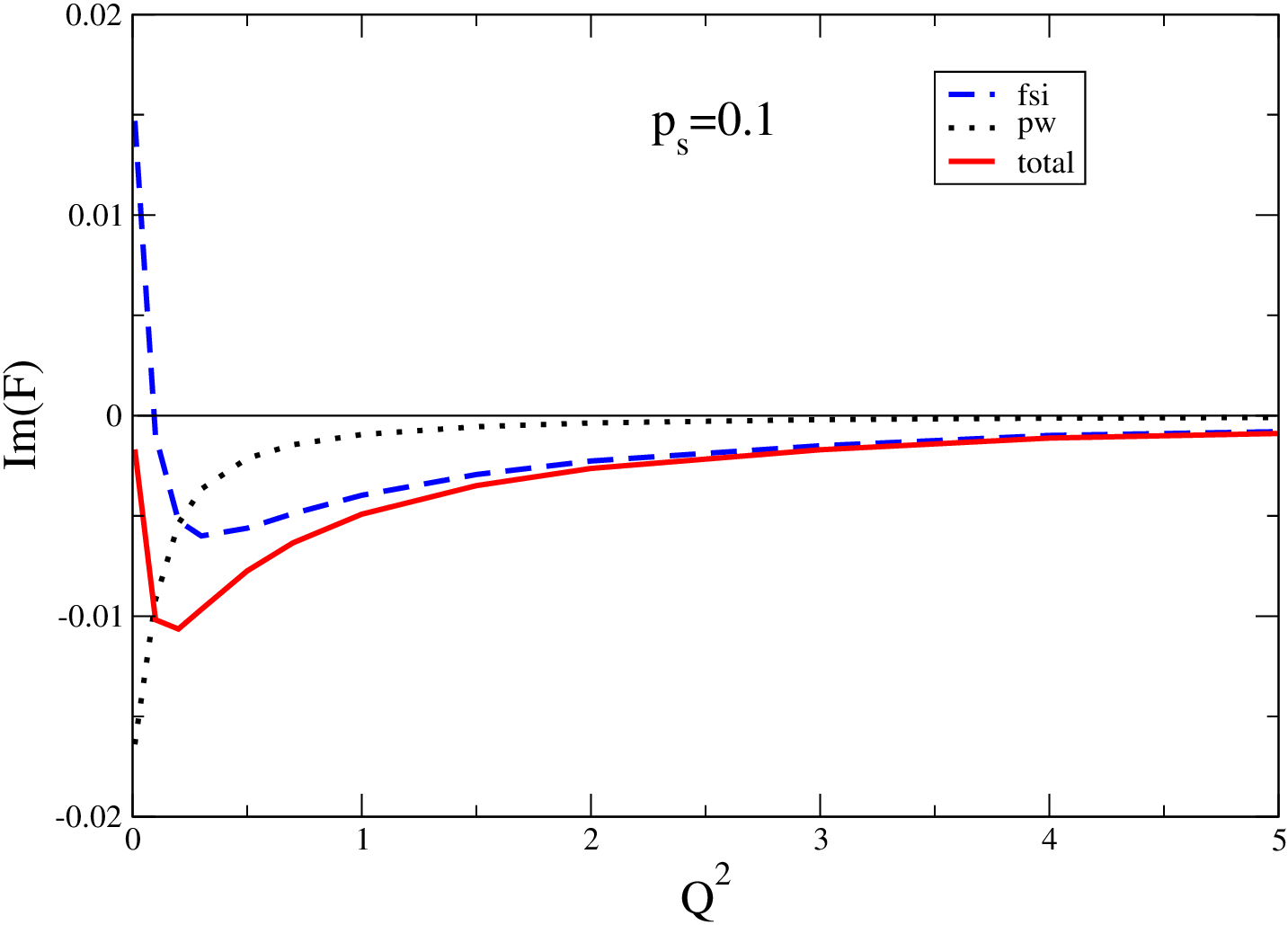}
\caption{
Transition EM form factor $F(Q^2)$ as a function of $Q^2$. Initial (bound) state corresponds to the binding energy $B=0.01\,m$; final (scattering) state corresponds to 
a relative momentum $p_s=0.1\,m$ (final state mass $M'=2.00998\,m$). FSI contribution is shown by the dashed curve and the PW one by a dotted curve. Full form factor  is shown by the solid curve. Left panel is the real part of form factor and right panel is the imaginary part.}
\label{ff}
\end{center}
\vspace{-1cm}
\end{figure}

We show in Fig. \ref{ff} the transition form factor for the  initial (bound) state  binding energy $B=0.01\,m$ (the initial state mass $M=1.99\,m$) and for the final (scattering) state  relative momentum $p_s=0.1\,m$ with corresponding final state mass values $M'=2\sqrt{m^2+p_s^2} \approx 2.00998\,m$. It is the sum of two contributions: final state interaction (FSI) and plane wave (PW).
In contrast to the elastic scattering, the inelastic transition form factor is complex. Its real and imaginary parts  as a function of $Q^2$ for $p_s=0.1\,m$ are shown in Fig. \ref{ff}, at left and right panels correspondingly.  One can see that at relatively small momentum transfer $Q^2 < 1$ both contributions -- FSI and PW -- are equally important and they considerably cancel each other.

The gauge invariance, i.e.  the conservation of the transition e.m. current $J\cdot q=0$, is verified in \cite{ck_fft} numerically.
It results from full cancellation between  the PW  and FSI contributions to  $J\cdot q$
which  takes place only if the bound and scattering state  amplitudes, and the e.m. current  are strictly consistent with each other.  This cancellation takes place numerically with rather high precision, that confirms the validity of our calculations.

The results shown above -- the off-shell amplitude, the phase shifts, the scattering length and e.m. form factors -- 
represent the exhaustive set of calculations for a two-body system. They put on the agenda the similar treatment, in the BS framework, of the realistic systems (nucleons and hadrons).


%
%
%

\end{document}